\documentstyle[12pt,worldsci]{article}
\catcode`@=11
\newtoks\@stequation

\def\mathletters{\refstepcounter{equation}%
  \edef\@savedequation{\the\c@equation}%
  \@stequation=\expandafter{\theequation}
  \edef\@savedtheequation{\the\@stequation}
  \edef\oldtheequation{\theequation}%
  \setcounter{equation}{0}%
  \def\theequation{\oldtheequation\alph{equation}}}

\def\endmathletters{%
  \setcounter{equation}{\@savedequation}%
  \@stequation=\expandafter{\@savedtheequation}%
  \edef\theequation{\the\@stequation}%
  \global\@ignoretrue}
\catcode`=12

\def\PL #1 #2 #3 {Phys. Lett.~{\bf#1} (#2) #3}
\def\NP #1 #2 #3 {Nucl. Phys.~{\bf#1} (#2) #3}
\def\ZP #1 #2 #3 {Z.~Phys.~{\bf#1} (#2) #3}
\def\PR #1 #2 #3 {Phys. Rev.~{\bf#1} (#2) #3}
\def\PRD #1 #2 #3 {Phys. Rev.~D {\bf#1} (#2) #3}
\def\PP #1 #2 #3 {Phys. Rep.~{\bf#1} (#2) #3}
\def\PRL #1 #2 #3 {Phys. Rev.~Lett.~{\bf#1} (#2) #3}
\def\etal {{\it et al}.}

\begin{document}
\font\fortssbx=cmssbx10 scaled \magstep2
\hbox to \hsize{
\hskip.5in \raise.1in\hbox{\fortssbx University of Wisconsin - Madison}
\hfill$\vcenter{\hbox{\bf MAD/PH/775}
            \hbox{July 1993}}$ }

\title{Low Energy Constraints and Anomalous Triple Gauge Boson
Couplings\footnote{Talk presented at the {\it Workshop on Physics and
Experiments with Linear $e^+e^-$ Colliders}, Waikoloa, Hawaii, April 26--30,
1993.}}
\author{Dieter Zeppenfeld\\
\it
Department of Physics, University of Wisconsin, Madison, WI 53706, USA}
\maketitle

\begin{center} ABSTRACT\\ [.1in]
\parbox{13cm}{\small
Low energy (1-loop) constraints on anomalous triple gauge boson vertices
(TGV's) are revisited and compared to the sensitivity achievable at LEP II and
at future linear $e^+e^-$ colliders. The analysis is performed within the
framework of an effective Lagrangian of gauge invariant dimension six operators
with the gauge bosons and a single Higgs doublet field as the low energy
degrees of freedom. The low energy data do not directly bound TGV's but they
provide strong constraints on models which lead to anomalous gauge boson
interactions in addition to other low energy effects. }
\end{center}

\section{Introduction}

Over the last four years $e^+e^-$ collision experiments at LEP and at the SLAC
linear collider have beautifully confirmed the predictions the Standard Model
(SM). At present experiment and theory generally agree at the 1\% level or
better in the determination of the vector boson couplings to the various
fermions\cite{rolandi}, which may rightly be considered a confirmation
of the gauge boson nature of the $W$ and the $Z$. Nevertheless
the most direct consequence of the $SU(2)\times U(1)$ gauge symmetry, the
nonabelian self-couplings of the $W$, $Z$, and photon, remain poorly
measured to date. Even if the underlying theory is
$SU(2)\times U(1)$ invariant, novel strong interactions in the
gauge boson--Higgs sector may lead to anomalous $WWZ$ and
$WW\gamma$ couplings.

One of the major reasons for raising the energy of the LEP collider above the
$W$-pair threshold is the systematic study and measurement of these triple
gauge boson vertices (TGV's) via the process
$e^+e^- \to W^+W^-$\cite{LEPWW,HPZH}. $W$ pair production together with
measurements of the single $W$ production cross section at a future linear
$e^+e^-$ or $e\gamma$ collider will provide us with an excellent measurement
of the three vector boson couplings\cite{singleW}. One can
quantify the sensitivity of all these experiments by parameterizing the most
general $WWV\, (V=Z,\gamma)$ vertex in terms of an effective Lagrangian
${\cal L}_{eff}^{WWV}$. Considering $C$ and $P$ even couplings only, it takes
the form\cite{HPZH}
\begin{equation}
i{\cal L}_{eff}^{WWV} = g_{WWV}\, \left( g_1^V(
W^{\dagger}_{\mu\nu}W^{\mu}-W^{\dagger\, \mu}W_{\mu\nu})V^{\nu} +
\kappa_V\,  W^{\dagger}_{\mu}W_{\nu}V^{\mu\nu} + {\lambda_V\over m_W^2}\,
W^{\dagger}_{\rho\mu}{W^{\mu}}_{\nu}V^{\nu\rho}\right)\; . \label{LeffWWV}
\end{equation}
Here the overall coupling constants are defined as $g_{WW\gamma}=e$ and
$g_{WWZ}= e \cot\theta_W$. Within the SM the couplings are given by
$g_1^Z = g_1^\gamma = \kappa_Z = \kappa_\gamma = 1,\; \lambda_Z =
\lambda_\gamma = 0$.

While the value of $g_1^\gamma$ is fixed by electromagnetic gauge invariance
(it is just the electric charge of the $W^+$) the other couplings may well
deviate from their SM values and need to be determined experimentally.
At LEP II one expects a sensitivity to deviations from the SM predictions of
$\Delta\kappa \approx \Delta\lambda \approx 0.1...0.2$ while the future
$e^+e^-$
linear colliders will push the precision of these measurements to the 1\%
level or below\cite{explin}.

\section{Effective Lagrangians}

The question arises whether the present high precision measurements at LEP
and at lower energies already give comparable constraints via 1-loop
corrections to $S$-matrix elements which involve the $WWV$ vertices. Many such
investigations have been performed in the past\cite{muf,suzuki,grifols,Hagi},
usually, however, in a framework which introduces the deviations from the SM
in such a way as to violate $SU(2)\times U(1)$ gauge-invariance. As a result
the 1-loop contributions from anomalous $WWV$ interactions
to oblique parameters like $\delta\rho$ or the $S,T,U$ parameters
of Peskin and Takeuchi\cite{STU} turn out to be quadratically or even
quartically divergent \cite{suzuki}. This in turn has lead to very stringent
bounds from existing low-energy data.

While the effective Lagrangian ${\cal L}_{WWV}$ of Eq.(\ref{LeffWWV}) is
general
enough for a discussion of weak boson pair production, low energy observables
are affected at the 1-loop level not only by the TGV's. One also expects
contributions from other new interactions which are induced by the new physics
simultaneously with anomalous values of $g_1$, $\kappa$, or $\lambda$.
In order to take such effects into account while avoiding an inflation of free
parameters, some simplifying assumptions are needed.

Given the excellent agreement of the measured fermion couplings with the SM
gauge theory predictions, I shall assume in the following that
\begin{enumerate}
\item[i)]
The $W$, $Z$, and photon are indeed the gauge bosons of a spontaneously broken
$SU(2)\times U(1)$ local symmetry.
\item[ii)]
New contributions to the gauge boson--fermion couplings can be neglected.
\item[iii)]
The low energy effects of the new interactions which are responsible for
anomalous $WWV$ couplings are described by an effective Lagrangian with the
$SU(2)\times U(1)$ gauge fields and the Higgs doublet field
as the low energy degrees of freedom:
\begin{eqnarray}\label{Leff}
{\cal L}_{eff} = \sum_i {f_i \over \Lambda^2}{\cal O}_i +
                 \sum_i {f_i^{(8)} \over \Lambda^4}{\cal O}_i^{(8)}+\,...\;\ .
\end{eqnarray}
\end{enumerate}

Here the scale $\Lambda$ may be identified with the typical mass of new
particles associated with the new physics.
Because of assumptions i) and ii) only gauge invariant operators ${\cal O}_i$
are allowed which can be constructed out of the Higgs field $\Phi$, covariant
derivatives of the Higgs field, $D_\mu \Phi$, and the field strength tensors
$W_{\mu\nu}$ and $B_{\mu\nu}$ of the $W$ and the $B$ gauge fields:
\begin{eqnarray}\label{Wmunu}
[D_\mu,D_\nu] = \hat{B}_{\mu\nu} + \hat{W}_{\mu\nu} =
                i\,{g'\over 2}\,B_{\mu\nu} +
                i\,g\,{\sigma^a\over 2}W^a_{\mu\nu}\; .
\end{eqnarray}

The use of a linear realization of the electroweak symmetry breaking in terms
of the Higgs doublet field $\Phi$ allows to discuss Higgs mass effects in the
following. It is general enough, though, since nonlinear realizations of the
symmetry breaking sector can be simulated by the $m_H\to \Lambda$ limit. As
has been emphasized by Burgess and London\cite{bl} the gauge invariance
assumption does not really provide any constraints on $e.g.$ anomalous TGV's
induced by ${\cal L}_{eff}$, since the phenomenological
Lagrangian ${\cal L}_{WWV}$ can be regarded as the unitary gauge version of
an explicitly $SU(2)\times U(1)$ invariant effective Lagrangian.  Constraints
arise when making one additional assumption:

\begin{enumerate}
\item[iv)]
The effective Lagrangian may be truncated at the dimension six level, $i.e.$
corrections of order $m_W^2/\Lambda^2$ or $v^2/\Lambda^2$ can be neglected in
the low energy effects.
\end{enumerate}

\noindent
This last assumption, while limiting the applicability of the subsequent
analysis somewhat, is general enough to elucidate the generic problems of low
energy constraints on the $WWV$ couplings, as we shall see later.

A complete list of $SU(2)\times U(1)$ invariant dimension six operators has
been given in Ref.~12 and has by now been employed in the analysis of
$WWV$ couplings by many authors\cite{deR,HISZ,hv,HISZnew,Einhorn,schildknecht}.
Using the SM equations of motion for the Higgs doublet field and identifying
operators which only differ by a total derivative, 11 independent operators
can be constructed at the dimension six level. Of these only 9 contribute to
four-fermion amplitudes up to 1-loop:
\begin{eqnarray}\label{operators}
{\cal L}_{eff} & = & \sum_{i=1}^9 {f_i\over \Lambda^2} {\cal O}_i =
{1\over \Lambda^2} \biggl( 
f_{\Phi,1}\; (D_\mu\Phi)^\dagger \Phi\; \Phi^\dagger (D^\mu \Phi)\;
+\; f_{BW}\; \Phi^\dagger\hat{B}_{\mu\nu}\hat{W}^{\mu\nu} \Phi\;
\nonumber \\
& + & f_{DW}\; Tr([D_\mu, \hat{W}_{\nu\rho}]\, [D^\mu,\hat{W}^{\nu\rho}])\;
-\; f_{DB}\; {g'^2\over 2}(\partial_\mu B_{\nu\rho})(\partial^\mu
B^{\nu\rho})\; \nonumber  \\
& + & f_B\; (D_\mu\Phi)^\dagger\hat{B}^{\mu\nu}(D_\nu\Phi)\;
+\; f_W\; (D_\mu\Phi)^\dagger\hat{W}^{\mu\nu}(D_\nu\Phi)\;
+\; f_{WWW}\; Tr[\hat{W}_{\mu\nu}\hat{W}^{\nu\rho}{\hat{W}_\rho}\,^\mu]
\nonumber \\
& + & \; f_{WW}\; \Phi^\dagger\hat{W}_{\mu\nu}\hat{W}^{\mu\nu}\Phi\;
+\; f_{BB}\; \Phi^\dagger\hat{B}_{\mu\nu}\hat{B}^{\mu\nu}\Phi
\biggr) \; ,
\end{eqnarray}

The first four operators, ${\cal O}_{\Phi,1}$, ${\cal O}_{BW}$,
${\cal O}_{DW}$, and ${\cal O}_{DB}$,
affect the gauge boson two-point functions at tree level\cite{gw} and as a
result the coefficients of these four operators are
severely constrained by present low energy data.

Of the remaining five, ${\cal O}_{WWW},\,\,{\cal O}_W$ and ${\cal O}_B$ give
rise to non-standard triple gauge boson couplings. Their presence in the
effective Lagrangian leads to deviations of the $WWV$ couplings from the SM,
namely\cite{deR,HISZnew}
\begin{mathletters}
\begin{eqnarray}\label{kapanom}
\kappa_\gamma = 1 + (f_B+f_W)\;{m_W^2\over 2\Lambda^2}\; , \;&&\;\;
\kappa_Z = 1 + \left(f_W-s^2(f_B+f_W)\right)\;
                 {m_Z^2\over 2\Lambda^2}\; , \\
g_1^Z = 1 + f_W\;{m_Z^2\over 2\Lambda^2} & = &
\kappa_Z + {s^2\over c^2}(\kappa_\gamma - 1) \; , \\
\lambda_\gamma = \lambda_Z & = & {3m_W^2g^2\over 2\Lambda^2}\; f_{WWW}
=\lambda\; ,\label{kapanome}
\end{eqnarray}
\end{mathletters}
with $s = {\rm sin}\theta_W$. As mentioned earlier, the correlations between
different anomalous $WWV$ couplings exhibited in the last two equations are
due to the truncation of the effective Lagrangian at the dimension six level
and do not hold any longer when dimension eight operators are
included\cite{HISZnew}. In addition to anomalous triple boson vertices the
operators of Eq.~(\ref{operators}) provide anomalous gauge boson Higgs
couplings, and these additional interactions play a very important role in
canceling the divergencies which have plagued earlier loop analyses.

\section{Low Energy Observables}

The effective Lagrangian of Eq.~(\ref{operators}) can now be used to
calculate loop corrections involving anomalous $WWV$ couplings. A range of
observables have been considered in the past. Classical examples are
$(g-2)_\mu$ \cite{muf,arzt}, and the $b\to s\gamma$ decay rate\cite{btosgamma}.
I will not discuss these
processes in detail here because of two reasons: i) the oblique corrections
give more stringent constraints on the 11 operators considered above and ii)
a complete discussion along the lines given below for the oblique corrections
necessitates the introduction of additional operators which involve
fermions\cite{arzt}. In the case of the anomalous magnetic moment of the muon
this is $e.g.$ the magnetic moment operator
\begin{equation}\label{ogm2mu}
{\cal O}_{g-2} = {m_\mu \over v} \bar L \sigma^{\mu\nu}
(a_W \hat W_{\mu\nu} + a_B \hat B_{\mu\nu} ) \Phi \mu_R\;,
\end{equation}
where $L$ denotes the left-handed $(\nu_\mu,\mu)$ doublet field.
These direct tree level contributions need to be considered together with the
loop corrections involving $WWV$ couplings in order to perform a
model-independent analysis of the low energy bounds and for absorbing the
divergencies of the loop integration\cite{arzt}.

These problems appear as well in the analysis of oblique corrections involving
the operators which were discussed in the previous section, and my discussion
here closely follows the one in Ref.~16. A vast amount of
experimental data can be understood as the measurement of 4-fermion $S$-matrix
elements. This includes the recent LEP
data, neutrino scattering experiments, atomic parity violation, $\mu$-decay,
and the $W$-mass measurement at hadron colliders. Since the data are now
sensitive to electroweak loop-corrections, these SM corrections must be
considered at the same time as the new physics contributions. After
correcting for SM box contributions and non-universal vertex corrections
(in particular the top mass dependence of the $Zb\bar b$-vertex) the
remaining SM contributions as well as all divergent new physics contributions
can be parameterized in a simple way. The four-fermion amplitudes
for massless external fermions are given by
\begin{eqnarray}\label{m4fermion}
{\cal M}(p_1,p_2,p_3,p_4) = I(q^2)\; J_\mu (p_1,p_2) J^\mu (p_3,p_4)\; .
\end{eqnarray}
Here the $J_\mu$ only depend on the wave functions of the external fermions
and the helicity dependent $I(q^2)$ are given by
\begin{eqnarray}\label{ICC}
I_{CC}(q^2) = {\bar g_W^2(q^2)/2 \over q^2-m_W^2+im_W\Gamma_W}
\end{eqnarray}
for CC amplitudes of left-handed fermions, while NC amplitudes may be
written as
\begin{equation}\label{INC}
I_{NC}(q^2) =  {\bar e^2(q^2)\over q^2}Q_{f_1}Q_{f_3} 
 +  {\bar g_Z^2(q^2)\over q^2-m_Z^2+im_Z\Gamma_Z}
\biggl( T_3^{f_1}-\bar s^2(q^2)Q_{f_1}\biggr)
\biggl( T_3^{f_3}-\bar s^2(q^2)Q_{f_3}\biggr)\; .
\end{equation}
$Q_{f_i}$ denotes the electric charge and $T_3^{f_i}$ the third component of
the weak isospin of fermion $f_i$.

The free parameters, which need to be determined by experiment, are the four
form-factors $\bar e^2(q^2)$, $\bar g_W^2(q^2)$, $\bar g_Z^2(q^2)$, and
$\bar s^2(q^2)$ and the $W$ and $Z$ mass. Three measurements are needed to
define the parameters of the SM, and these may be chosen as $m_Z$,
the Fermi constant $G_F \propto \bar g_W^2(0)/m_W^2$, and
$\alpha = \bar e^2(0)/4\pi$. Only after these values have been fixed can the
remaining data be used to place constraints on new physics contributions.

An analysis of the available data has recently been performed by Hagiwara et
al.\cite{kaorusm} For $m_t = 140$~GeV the LEP and SLC data can be summarized
in terms of
\begin{eqnarray}\label{gzmzdata}
\bar g_Z^2(m_Z^2)  =  0.5524 \pm 0.0017  \;,\;\;\;\;\;\;\;\;\
\bar s^2(m_Z^2)  =  0.2319 \pm 0.0011  \;.
\end{eqnarray}
A slight top mass dependence of the extracted results is negligible compared
to the errors.
In a similar fashion the low-energy data on neutrino scattering and atomic
parity violation determine the same form-factors at zero momentum transfer:
\begin{eqnarray} \label{gz0data}
\bar g_Z^2(0)  =  0.5462 \pm 0.0035  \;,\;\;\;\;\;\;\;\;\
\bar s^2(0)  =  0.2359 \pm 0.0048  \;.
\end{eqnarray}
Finally, the $W$-mass measurement at hadron colliders together with the input
value of $G_F$ can be translated into a measurement of $\bar g_W^2(0)$:
\begin{eqnarray}
\bar g_W^2(0) = 0.4217 \pm 0.0027  \;. \label{gw0data}
\end{eqnarray}

These five measurements are closely related to other formulations of the
oblique corrections, like the S,T,and U parameters of Peskin and
Takeuchi\cite{STU}. S and T, for example, are given by\cite{HISZnew}
\begin{mathletters}
\begin{eqnarray}
S &=& {4\bar s^2(m^2_Z)\bar c^2(m^2_Z)\over
\bar\alpha(m^2_Z)_{\rm SM}} - {16\pi\over \bar g^2_Z(0)} \;, \label{S} \\
1-\alpha T &=& {1\over\bar\rho} \equiv {\bar g^2_W(0)\over \bar g^2_Z(0)}\,
{m^2_Z\over m^2_W}\;. \label{T}
\end{eqnarray}
\end{mathletters}

The new feature here is the inclusion of the $q^2$ dependence of the
form-factors\cite{HISZnew,kaorusm,maksymyk}. Indeed, new physics contributions
like the operators ${\cal O}_{DW}$ or ${\cal O}_{DB}$ do lead to a nontrivial
$q^2$ dependence of the form-factors in Eq.~(\ref{INC}), and the more general
analysis is needed to constrain these operators. Low energy bounds are
obtained by fitting
\begin{mathletters}
\begin{eqnarray}
S &=& S_{SM}(m_t,m_H) + \Delta S\; , \\
T &=& T_{SM}(m_t,m_H) + \Delta T\;\;\; {\rm etc.}
\end{eqnarray}
\end{mathletters}
to the data. Here the SM contributions ($S_{SM}$ etc.) introduce a significant
dependence on the as yet unknown values of the Higgs and the top quarks masses.

The four operators ${\cal O}_{DW}$, ${\cal O}_{DB}$, ${\cal O}_{BW}$, and
${\cal O}_{\Phi,1}$, contribute already at tree level,
\begin{mathletters}
\begin{eqnarray}\label{resff}
\Delta\delta\rho & = & \alpha \Delta T =
- {v^2\over 2\Lambda^2}\; f_{\Phi,1}\; , \\
\Delta S & = & - 32\pi s^2{m_W^2\over \Lambda^2}\; (f_{DW}+f_{DB})
               -4\pi{v^2\over\Lambda^2}f_{BW}\; ,
\end{eqnarray}
\end{mathletters}
with similar results for the other form-factors. Fitting these to the five
data points one obtains measurements of the coefficients of the operators in
the effective Lagrangian,
\begin{mathletters}
\begin{eqnarray}\label{valfi}
f_{DW}/\Lambda^2 & = & (0.56 \pm 0.79)\; {\rm TeV}^{-2} \; ,\\
f_{DB}/\Lambda^2 & = & (-8.0 \pm 11.9)\; {\rm TeV}^{-2} \; ,\\
f_{BW}/\Lambda^2 & = & (1.9 \pm 2.9)\; {\rm TeV}^{-2} \; ,\\
f_{\Phi,1}/\Lambda^2 & = & (0.11 \pm 0.20)\; {\rm TeV}^{-2} \; ,
\end{eqnarray}
\end{mathletters}
for $m_H = 200$~GeV and $m_t = 140$~GeV. While the central values depend on
the choice of $m_t$ and $m_H$, the quoted errors are unaffected. There are
strong correlations between the coefficients of the dimension six operators,
however, in particular between $f_{DB}$, $f_{BW}$ and $f_{\Phi,1}$.

While the contributions of these four operators are already constrained at the
tree level, the remaining five, which include the anomalous $WWV$ couplings,
only contribute at the 1-loop level to the oblique correction parameters.
Contributions to the four-fermion amplitudes arise via the corrections to the
gauge boson self-energies and also to the gauge boson--fermion vertices. In
fact both need to be included to preserve gauge invariance. The complete
calculation of the logarithmically enhanced contributions was performed in
Ref.~16, partial results can be found in Refs.~13--15.

At intermediate steps of the calculation one still encounters quadratic
divergencies due to the insertion of dimension six operators in the loops.
These
are all absorbed, however, into the renormalization of the SM parameters
$m_Z$, $G_F$ and $\alpha$. All remaining logarithmically divergent terms are
found to be renormalizations of the four operators which already contributed
at tree level. Neglecting all terms which are not logarithmically enhanced,
the leading effects are given by replacing $f_{DW}$ etc. in Eq.~(15)
by the renormalized quantities
\begin{mathletters}
\begin{eqnarray}\label{firen}
f_{DW}^r & = & f_{DW} -
        {1\over 192\pi^2}f_W\; {\rm log}{\Lambda^2\over \mu^2}\; , \\
f_{DB}^r & = & f_{DB} -
        {1\over 192\pi^2}f_B\; {\rm log}{\Lambda^2\over \mu^2}\; , \\
f_{BW}^r & = & f_{BW} + {\alpha\over 32\pi s^2}{\rm log}{\Lambda^2\over\mu^2}
\biggl(f_B({20\over3}+{7\over 3c^2}+{m_H^2\over m_W^2})
-f_W(4+{1\over c^2}-{m_H^2\over m_W^2})    \nonumber  \\      &&
+12g^2f_{WWW} - 8(f_{WW}+{s^2\over c^2}f_{BB}) \biggr)  \\
f_{\Phi,1}^r & = & f_{\Phi,1} + {3\alpha\over 8\pi c^2}
{\rm log}{\Lambda^2\over\mu^2}
\left(f_B{m_H^2\over v^2}+{3m_W^2\over v^2}(f_B+f_W)\right)\; .
\end{eqnarray}
\end{mathletters}
After renormalization of the SM parameters and of the operators which
contribute at tree level, the remaining corrections to four-fermion amplitudes
are finite. The ${\rm log}{\Lambda^2\over\mu^2}$ terms in Eq.~(17)
describe mixing of the operators between the new physics scale $\Lambda$ and
the weak boson mass scale $\mu=m_W$.

The quadratic divergencies observed in earlier work are cancelled by Higgs
contributions to the vacuum polarization of the $W$ and the $Z$: $SU(2)\times
U(1)$ gauge invariance and the use of a linear realization for the symmetry
breaking sector relates TGV's to anomalous Higgs-gauge boson
interactions. Gauge invariance guarantees the cancellation of all quadratic
divergencies  between gauge boson and Higgs contributions\cite{deR}. A trace
of the quadratic divergencies is preserved in the $m_H^2$ terms in the results
of Eqs.~(17): the Higgs graphs give rise to
$-\Lambda^2+m_H^2{\rm log}\Lambda$ terms. By including Higgs exchange we have
therefore replaced quadratic divergencies by $m_H^2$ terms. In the limit
$m_H\to \Lambda$ the quadratic divergencies are recovered.

\section{Low Energy Bounds on Anomalous $WWV$ Couplings?}

We have seen that all divergent 1-loop contributions involving TGV's are just
renormalizations of the coefficients of some other, independent, operators.
Hence these $\Lambda^2$ or log$\;\Lambda$ terms cannot be used for a direct
measurement of the $WWV$ couplings without making assumptions on the absence of
cancellations between tree level and 1-loop contributions. Even including the
finite corrections involving $WWV$ couplings the five data points of
Eqs.~(\ref{gzmzdata}--\ref{gw0data}) are barely sufficient to limit the four
tree level coefficients $f^r_{DW}/\Lambda^2$, $f^r_{DB}/\Lambda^2$,
$f^r_{BW}/\Lambda^2$, and $f^r_{\Phi,1}/\Lambda^2$ in addition to the SM top
quark and Higgs boson mass dependences. Without additional assumptions the
$WWV$ couplings remain unconstrained by the present low-energy data.

One may assume, for example, that $f_{BW}$ vanishes at the scale
$\Lambda =1$~TeV and that the main contribution to $f_{BW}^r$ arises at 1-loop
from $\lambda= 3\,g^2\, f_{WWW}\;m_W^2/2\Lambda^2$. Eq.~(16c) then
translates into
\begin{eqnarray}\label{lambdabound}
\lambda  = 0.89 \pm 1.35 \; ,
\end{eqnarray}
a constraint which is comparable to present hadron collider bounds\cite{UA2}.

The more traditional analysis of low energy bounds assumes that only one
coupling differs from its SM value, hence that no cancellations occur between
the contributions from different couplings. In the context of our 11
dimension-six operators this corresponds to considering the three cases
$f_B\neq 0$, $f_W \neq 0$, and $f_{WWW}\neq 0$ with the coefficients of the
remaining 10 operators vanishing. Choosing $m_t=140$~GeV and $m_H=200$~GeV one
finds\cite{HISZnew}
\begin{mathletters}
\begin{eqnarray}\label{constraintf}
\kappa_\gamma = 1 + f_B\; {m_W^2\over 2\Lambda^2} &=& 1.04\pm 0.06\; \;\;\;
{\rm for} \;\;\; f_B\neq 0\;, \\
\kappa_\gamma = 1 + f_W\; {m_W^2\over 2\Lambda^2} &=& 1.01\pm 0.09\; \;\;\;
{\rm for} \;\;\; f_W\neq 0\;, \\
\lambda_\gamma =  {3m_W^2g^2\over 2\Lambda^2}\; f_{WWW} &=& 0.03\pm 0.16\;
\;\;\; {\rm for} \;\;\; f_{WWW}\neq 0\;.
\end{eqnarray}
\end{mathletters}

A more stringent assumption has been proposed by De~R\'ujula et al.\cite{deR}
There are no obvious symmetries which distinguish the tree level operators
${\cal O}_{BW}$, ${\cal O}_{DW}$, ${\cal O}_{DB}$, and ${\cal O}_{\Phi,1}$
from the remaining seven. For a generic model of the underlying dynamics one
may hence expect $e.g.$ $|f_B+f_W| \approx |f_{BW}|$ which with
the result of Eq.~(16c) implies
$|\kappa_\gamma -1|= |f_B+f_W|\; {m_W^2\over 2\Lambda^2} < 0.02$ at "90\% CL",
a value too small to be observable in $W^+W^-$ production at LEP, but still in
the interesting range for future linear colliders.

As Einhorn et al.\cite{Einhorn} have argued, one should perhaps not expect
anomalous couplings which are larger than these most stringent bounds.
In extensions of the SM anomalous $WWV$ couplings arise from heavy particle
loops (of mass $M$) with three external gauge bosons attached. Because of
the universal factor $1/16\pi^2$ for loop integrals one should expect
\begin{eqnarray}\label{einhornbound}
{f_i\over \Lambda^2} = {1\over 16\pi^2}{c_f\over M^2}\;,
\end{eqnarray}
and hence $|f_i|m_W^2/\Lambda^2 < 10^{-3}$ even for masses as low as
$M=250$~GeV, unless the counting factor $c_f$ is substantially larger than
unity, e.g. due to higher isospin multiplets or because of large multiplicities
of the heavy particles.

The last two arguments indicate the difficulty of constructing realistic
models which would predict large anomalous $WWV$ couplings. One must clearly
state, however, that there is no proof that large anomalous couplings are
ruled out, and naturalness arguments may well prove erroneous. The eleven
dimension-six operators of Section~II are independent and must therefore be
constrained individually by experiment. For the $WWV$ couplings, $W^+W^-$
production and single $W$ production at future linear colliders are the ideal
way to achieve this goal.

\newpage

\section*{Acknowledgements}
I would like to thank K.~Hagiwara, S.~Ishihara, and R.~Szalapski for an
enjoyable collaboration which led to many of the results presented in this
talk.
This research was supported in part by the University of Wisconsin Research
Committee with funds granted by the Wisconsin Alumni Research Foundation,
by the U.~S.~Department of Energy under contract No.~DE-AC02-76ER00881,
and by the Texas National Research Laboratory Commission under Grants
No.~RGFY93-221 and FCFY92-12.

\end{document}